\documentclass[pra,twocolumn,showpacs,nobibnotes,floatfix]{revtex4}

\usepackage{graphicx,eucal,amsfonts,epsfig}
\input{diagramsext}
\diagramstyle[height=8mm,width=1.4cm]
\newarrow{Bond}	----{>}
\newcommand{\opr}[1]{\mathrm{#1}}

\begin{document}

\title{Reliable Final Computational Results from Faulty Quantum Computation}
\author{Gerald Gilbert, Michael Hamrick, Yaakov S. Weinstein}
\email{{ggilbert, mhamrick, weinstein}@mitre.org}
\affiliation{Quantum Information Science Group \\{\sc  Mitre} \\
260 Industrial Way West, Eatontown, NJ 07724 USA}

\begin{abstract}

In this paper we extend both standard fault tolerance theory and Kitaev's model for quantum computation,  
combining them so as to yield quantitative results 
that reveal the interplay between the 
two.  Our analysis establishes a methodology that allows us to quantitatively 
determine design parameters for a quantum 
computer, the values of which ensure 
that an overall computation of interest yields a correct {\em final result} with some prescribed probability of 
success, as opposed to merely ensuring that the desired {\em final quantum state} 
is obtained. 
As a specific example of the practical application of our approach, 
we explicitly calculate the number of levels of error correction 
concatenation needed to achieve a correct final result for the overall computation with 
some prescribed success probability.  
Since our methodology allows one to determine parameters required in order to achieve the 
correct final result for the overall quantum computation, as opposed to merely ensuring 
that the desired final quantum state is produced, our method enables the determination 
of {\em complete} quantum computational resource requirements associated to the actual solution 
of practical problems.  

\end{abstract}

   
\maketitle
\section{Introduction}

\begin{figure}
\begin{center}
\includegraphics[width = 8cm]{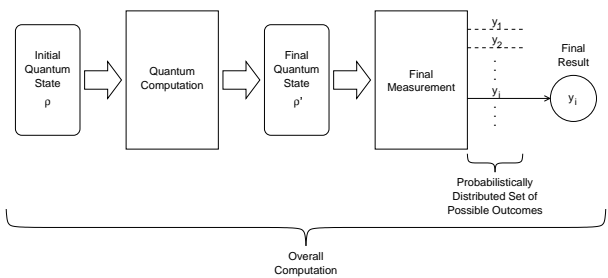}
\caption{\label{basicdefs}
Basic definitions.}
\end{center}
\end{figure}

The purpose of any computation, whether implemented by a quantum computer or a classical computer, is to compute the value of a function for specified values of the variables on which the function depends.  In the case of a quantum computer, the final result is obtained 
from the outcome of a final measurement performed on the final quantum 
state produced by the quantum computation.   
This is illustrated schematically in Figure \ref{basicdefs}, which also serves to fix 
the terminology we shall use in this paper to describe the quantum computation and the final 
measurement. In particular, in this paper, 
the term ``quantum computation" 
very specifically refers solely to the dynamical evolution 
of the qubits from some initial quantum state to some final quantum 
state.  Our use of this term does {\em not} include the 
subsequent, final measurement from which the final result of the overall 
computation is obtained.  We thus distinguish between the {\em final quantum state} of the 
{\em quantum computation} and the {\em final result} 
of the {\em overall computation}, of which the quantum computation is merely a part.

Since the final measurement produces, in general, a probabilistically distributed 
set of outcomes, the question arises as to whether or not reliable 
{\em final} results for the overall computation can be obtained. 
An affirmative answer to this question is necessary in order 
that algorithms such as Shor's algorithm \cite{Shor} can be successfully applied 
to practical problems. A partial answer to this question follows from
the quantum computational model formalized by Kitaev~(\cite{kitaev1997}, \S4.1). 
Kitaev's model identifies a bound, $p$, on the probability that the final result 
of the overall computation 
is incorrect due to the indeterminism of the final measurement. It follows that 
reliable overall computation 
can be achieved for sufficiently small $p$.  
However, Kitaev's model assumes that the quantum computation that 
produces the final quantum state (on which the final measurement is performed 
in order to give the final result of the overall computation) is perfectly 
implemented, with no 
errors. In other words, those problems that are 
within the purview of fault tolerance analysis, namely, the occurrence 
of 
errors in the {\em quantum} computation, and the effectiveness of error correction 
techniques in reducing the effect of such errors on the final {\em quantum} state, 
are not addressed in Kitaev's model.  
There are thus two sources of error that may affect the final result 
of the overall computation: (1) errors which arise in the course 
of the quantum computation, and (2) errors due to the indeterminacy intrinsic to the final 
measurement which follows the quantum computation~\cite{f1}.
Fault tolerance theory  ~\cite{Preskill,Lidar,book} addresses the first type of 
error, but does not consider the second.  Kitaev's model addresses the second type of error, but 
not the first.  
In the present paper, we extend both fault tolerance theory and Kitaev's model 
so as to take into account 
the combined effects of both sources of error, including the interplay between them.
  
%
%

In order to quantify the connection between fault tolerance theory and Kitaev's 
model, we utilize a measure of the difference between the desired final quantum state 
(that would arise in the case of perfect quantum computation) 
and the actual final quantum state (that is produced in an actual quantum computation realized by a practical device).  We define this measure in terms of a suitable norm 
of the difference between the two aforementioned 
states.  We refer to this measure as the ``implementation 
inaccuracy."  We then introduce a fundamental inequality, to which we refer as the 
Quantum Computer Condition (QCC), which requires 
that the implementation inaccuracy be less than some prescribed 
bound.  This fundamental inequality 
furnishes a criterion for successful fault tolerant quantum computation.  We shall see that the 
use of this criterion allows the conditions that ensure 
fault tolerance to be incorporated into Kitaev's model in 
a remarkably straightforward way.  

Our results provide a quantitative relationship between fault tolerance 
theoretic constraints on the one hand, 
and the accuracy of the final result of the overall computation on the other.  
Consequently, our results can be applied 
to explicitly and directly determine requirements on practical fault 
tolerant design parameters in order to achieve a specified accuracy for the {\em final result} of the overall computation. This goes 
beyond standard fault tolerance theory, which 
only addresses the accuracy with which the quantum computation realizes the desired 
{\em final quantum state}, and does not directly determine the accuracy of the final result of the overall computation.   
As an example of the application of our result, we show how to 
explicitly calculate the number of levels of 
concatenation of error correction required to directly achieve a specified probability 
that the overall computation produces the correct {\em final result}, and not merely the desired 
final quantum state. 

This paper is organized as follows.  
In Section 2 we review the Kitaev model and re-express it in a form that is convenient for 
our analysis.  In Section 3 we introduce the implementation inaccuracy and rigorously 
define the fundamental inequality known as the QCC.  In Section 4 we derive the constraints 
implied by combining the QCC with the Kitaev model.  As an example of the practical utility of 
our general result, we explicitly calculate the number of levels of error correction 
concatenation needed to achieve a correct final result for the overall computation with 
some prescribed success probability. In Section 5 we present our conclusions.

\section{The Kitaev Model}

We begin by reviewing the Kitaev model~\cite{kitaev1997}, which 
can be depicted by the following diagram:
\begin{equation} \label{commutative-diagram}
\begin{diagram}
H_\mathrm{logical}  & \rBond^{U} & H_\mathrm{logical} \\
\uBond<{\opr{I_{\mathbf{pure}}}} & p  &  \dBond>{\opr{O_{\mathbf{pure}}}} \\
X & \rBond_{F} & Y \end{diagram}~.
\end{equation}
This diagram expresses the fact that the output of the quantum computation 
$U$ is intended to be used in computing the function $F$,  where $F:X \rightarrow Y$ 
is an instance of the overall computational problem. Here ${\opr{I_{\mathbf{pure}}}}$ is an 
initialization map, which maps the input space $X$ into pure states of the 
logical Hilbert space $H_\mathrm{logical}$, and $\opr{O_{\mathbf{pure}}}$ is the corresponding readout map, which maps the output of the operation $U$ onto the output space $Y$. In general $\opr{O_{\mathbf{pure}}}$ is a measurement given by a projection-valued measure (or more generally a POVM) $\{\opr{E}_y\}_{y \in Y}$. The symbol in the center of the diagram denotes the probabilistic inaccuracy, $p$, associated to the output of the overall computation. If $p=0$, the computation always produces the 
correct result, and the diagram commutes.  Thus, the quantity $1-p$ furnishes a lower bound on the probability of the success of the overall computation.  We note that $p$ may be non-vanishing, even if the quantum 
algorithm is perfectly implemented, due to the fact that the measurement of the final quantum 
state produced by the quantum computation is necessarily quantum probabilistically distributed. Given an input $x \in X$, 
the final result of the overall computation is distributed according the probability
measure on $Y$ as follows: the probability $\opr{Pr}_x(y)$ of a
singleton $y \in Y$ is $\langle \opr{I}_{\mathbf{pure}}(x) | U^\dag  \opr{E}_y U 
| \opr{I}_{\mathbf{pure}}(x) \rangle$. Then, the {\em near commutativity} of
Diagram (\ref{commutative-diagram})
means that for each $x \in X$, the probability
measure $\opr{Pr}_x(y)$ is concentrated at $y = F(x)$, so that $F(x)$ is the final result 
of the 
overall computation with high probability.  Thus, Kitaev's formulation 
\cite{kitaev1997} requires that the Diagram~(\ref{commutative-diagram}) be nearly 
commutative in a probabilistic sense, that is 
\begin{equation}
\label{pure_eqn}
\langle \opr{I}_{\mathbf{pure}}(x) | U^\dag  \opr{E}_{F(x)} U  
          | \opr{I}_{\mathbf{pure}}(x) \rangle > 1-p~.
\end{equation}
In other words, the measurement of the final quantum state gives the correct final result for the overall computation 
with probability greater than $1-p$.  If $p$ is sufficiently small, {\em e.g.}, $p < 1/2$ in the case of 
$Y$ binary, a majority vote algorithm 
will successfully identify the correct outcome $y=F(x)$.  

We wish to extend the Kitaev model to apply to the scenario in which errors may occur in the implementation of the quantum computation. As a preliminary step, we re-express the Kitaev model in a more convenient form for our purposes by generalizing Diagram (\ref{commutative-diagram}) and eq.(\ref{pure_eqn}) through the  replacement of wavefunctions with density matrices. This allows us to discuss the effect of errors that transform pure states into mixed states.  
The generalized diagram is 
\begin{equation} \label{commutative-diagram_mixed}
\begin{diagram}
{\mathbf{T}}(H_\mathrm{logical})  & \rBond^{G} & {\mathbf{T}}(H_\mathrm{logical}) \\
\uBond<{\opr{I_{\mathbf{dm}}}} & p  &  \dBond>{\opr{O_\mathbf{dm}}} \\
X & \rBond_{F} & Y \end{diagram}~,
\end{equation}
where the action of the map $G$ is given by $G: \rho \mapsto U \rho U^\dagger$,
$\rho$ is the density matrix representing the initial state of the quantum computer,
the action of the map $\opr{I_\mathbf{dm}}$ is given by
$\opr{I_\mathbf{dm}}: x \mapsto | \opr{I}_{\mathbf{pure}}(x) \rangle \langle \opr{I}_{\mathbf{pure}}(x) |$, 
$\opr{O}_{\mathbf{dm}}$ is the measurement corresponding to $\opr{O}_{\mathbf{pure}}$, 
except that $\opr{O}_{\mathbf{dm}}$ acts on density matrices, and $\mathbf{T}(H)$ 
is the Banach space of trace class operators on a given Hilbert space, $H$. 
The probability $\opr{Pr}_x(y)$ of a
singleton $y \in Y$ is now $\opr{tr}(\sqrt{\opr{E}_y} U \opr{I}_\mathbf{dm}(x) U^\dagger
\sqrt{\opr{E}_y})$. The near commutativity of Diagram (\ref{commutative-diagram_mixed}) 
means that 
\begin{eqnarray}\label{approximation-cond}
\opr{tr}\left(\opr{O}_{\mathbf{dm}}\left(G\left(\opr{I}_{\mathbf{dm}}
\left(x\right)\right)\right)\right) \equiv \;\;\;\;\;\;\;\;\;\;\;\;\;\;\;\;\;\;\;\; & \nonumber\\
\opr{tr}\left(\sqrt{\opr{E}_{F(x)}} U \opr{I}_\mathbf{dm}(x) U^\dagger \sqrt{\opr{E}_{F(x)}}~\right) & > & 1 - p
\end{eqnarray}
for all $x \in X$. Eq.(\ref{approximation-cond}) generalizes eq.(\ref{pure_eqn}) to states 
described by density matrices.  

\section{Inclusion of Residual Errors and the QCC}

We now extend Kitaev's model by allowing for the inevitable survival of residual errors in 
any realistic implementation of a quantum computation, even upon successful application of fault-tolerance techniques. In other words, we will extend the 
analysis to include fault-tolerant operation, but in such a way as to ensure a prescribed 
success probability to achieve the correct final result of the overall computation.

We consider a function $F$, and an implementation
of an overall computation (including a quantum computation followed by a final measurement) 
that is intended to calculate $F$ for some value of the 
input. The relationship between the {\em ideal}, error-free quantum computation,
$G(\rho) = U\rho U^{\dag}$, defined on the logical Hilbert space, $H_\mathrm{logical}$,  
and the {\em actual} ~dynamical map, $P$, implemented by the physical 
quantum computer on the computational Hilbert space, $H_\mathrm{comp}$, can be 
described by the following diagram:
\begin{equation} \label{nearly-commutative-diagram}
\begin{diagram}
{\mathbf{T}}(H_\mathrm{comp})  & \rBond^{P} & {\mathbf{T}}(H_\mathrm{comp}) \\
\uBond<{\mathcal{M}_{{\{\mathrm{l}\rightarrow\mathrm{c}\}}}} & \alpha  &  
\dBond>{\mathcal{M}_{{\{\mathrm{c}\rightarrow\mathrm{l}\}}}} \\
{\mathbf{T}}(H_\mathrm{logical}) & \rBond_{G} & 
{\mathbf{T}}(H_\mathrm{logical}) \end{diagram}~,
\end{equation}
where the superoperators $\mathcal{M}_{\{\mathrm{l}\rightarrow\mathrm{c}\}}$ and 
$\mathcal{M}_{\{\mathrm{c}\rightarrow\mathrm{l}\}}$ are {\em linking maps} that 
are mathematically required to relate states in the logical space 
$H_\mathrm{logical}$ to states in the computational space $H_\mathrm{comp}$. 
This is because, in general, $H_\mathrm{logical}\neq H_\mathrm{comp}$ \cite{linkingmap}.

We formally express the content of Diagram 
(\ref{nearly-commutative-diagram}) {\em via} a relation between an idealized ($G$)
and actual ($P$) quantum computation:
\begin{equation}
\label{QC}
\|\mathcal{M}_{\{\mathrm{c}\rightarrow\mathrm{l}\}}(P\cdot(\mathcal{M}_{\{\mathrm{l}\rightarrow\mathrm{c}\}}(\rho))) -
G(\rho )\|_1 \leq \alpha ~,
\end{equation}
where $\| \cdot\|_1$ signifies the Schatten $1$-norm. The quantity on the 
left-hand-side of this inequality furnishes a measure of the inaccuracy of the 
actual implementation of $G$ by $P$. It tells us how well a practical quantum 
computing device actually implements an ideally defined quantum computation. 
We will refer to the left-hand-side of this inequality as the {\em implementation inaccuracy} of the quantum computation. The quantity, $\alpha$, on the right-hand-side of this inequality is a bound on the implementation inaccuracy. The value of $\alpha$ is {\em prescribed} as a requirement on the performance of the quantum computation. Thus, the diagram states that $P$ can implement the idealized, perfect quantum computation, $G$, with an inaccuracy no greater than $\alpha$. We refer to the entire inequality as the {\em Quantum Computer Condition} (QCC).

Thus, we have that Diagram (\ref{commutative-diagram_mixed}) connects the ideal quantum computation, 
$G$, to the calculation of $F$, and, separately, we have that 
Diagram (\ref{nearly-commutative-diagram}) 
connects the ideal quantum computation, $G$, to the actual implementation of the 
quantum computation, $P$, that is realized by a practical physical 
device. To establish the utility of practical quantum computers, what is needed now is a relation that 
connects the actual, device-implemented quantum computation, $P$, 
to the intended calculation of 
$F$. By analogy with eq.(\ref{approximation-cond}), we seek an expression of the form 
\begin{widetext}
\begin{equation}\label{des-approximation-cond}
\opr{tr}\left(\opr{O}_{\mathbf{dm}}\left(P^{\mathcal M}\left(\opr{I}_{\mathbf{dm}}
      \left(x\right)\right)\right)\right) \equiv
\opr{tr}\left(\sqrt{\opr{E}_{F(x)}} P^{\mathcal M} \left( \opr{I}_\mathbf{dm}(x) \right) 
\sqrt{\opr{E}_{F(x)}}~\right) >
1 - p^\prime~,
\end{equation}
for some $p^\prime$, where $P^\mathcal{M} \equiv \mathcal{M}_{\{\mathrm{c}\rightarrow\mathrm{l}\}} 
P \mathcal{M}_{\{\mathrm{l}\rightarrow\mathrm{c}\}}$.  
This is established with the following 
calculation:
\begin{eqnarray}
\opr{tr}{\Big (}{\opr{O_\mathbf{dm}}}(P^{\mathcal M}(\opr{I}_\mathbf{dm}(x))){\Big )} & = & 
\opr{tr} \left(\sqrt{\opr{E}_{F(x)}}  P^{\mathcal M}(\opr{I}_\mathbf{dm}(x)) 
\sqrt{\opr{E}_{F(x)}}~\right)\nonumber  \\
& = & \opr{tr}\left(\sqrt{\opr{E}_{F(x)}}\bigl\{ P^{\mathcal
M}(\opr{I}_\mathbf{dm}(x)) - U \opr{I}_\mathbf{dm}(x) U^\dagger \bigr\}
\sqrt{\opr{E}_{F(x)}}~\right)\nonumber \\
& + & \opr{tr}\left(\sqrt{\opr{E}_{F(x)}} U \opr{I}_\mathbf{dm}(x) U^\dagger 
\sqrt{\opr{E}_{F(x)}}~\right)\nonumber \\
& > & -\alpha + 1 - p,
\end{eqnarray}
where we have used the property of the Schatten 1-norm, 
\begin{eqnarray}
{\Big\vert} \opr{tr}\left(\sqrt{\opr{E}_{F(x)}}\bigl\{ P^{\mathcal M}
   (\opr{I}_\mathbf{dm}(x)) - U \opr{I}_\mathbf{dm}(x) U^\dagger \bigr\}
   \sqrt{\opr{E}_{F(x)}}~\right)\!\!{\Big\vert} \leq 
\|  P^{\mathcal M}
   (\opr{I}_\mathbf{dm}(x)) - U \opr{I}_\mathbf{dm}(x) U^\dagger   \|_1~,
\end{eqnarray}
and the condition 
\begin{equation}
\|  P^{\mathcal M}
   (\opr{I}_\mathbf{dm}(x)) - U \opr{I}_\mathbf{dm}(x) U^\dagger   \|_1 =
\|\mathcal{M}_{\{\mathrm{c}\rightarrow\mathrm{l}\}}(P\cdot(\mathcal{M}_{\{\mathrm{l}\rightarrow\mathrm{c}\}}(\opr{I}_\mathbf{dm}(x)))) -
G(\opr{I}_\mathbf{dm}(x))\|_1 \leq \alpha~.  
\end{equation}
\end{widetext}
The physical meaning of this result can be transparently expressed by combining the 
content of Diagrams (\ref{commutative-diagram_mixed}) and (\ref{nearly-commutative-diagram}) 
to yield
%
%
\begin{equation} \label{nearly-commutative-diagram-1}
\begin{diagram}
{\mathbf{T}}(H_\mathrm{comp})  & \rBond^{P} & {\mathbf{T}}(H_\mathrm{comp}) \\
\uBond<{\opr{{\tilde I}_\mathbf{dm}}} & \alpha + p  &  \dBond>{\opr{{\tilde O}_\mathbf{dm}}} \\
X & \rBond_{F} & 
Y \end{diagram}~.
\end{equation}
This diagram is nearly commutative in the sense that
\begin{eqnarray}\label{approximation-cond-for-devices}
\opr{tr}\left(\opr{O}_{\mathbf{dm}}\left(P^{\mathcal M}\left(\opr{I}_{\mathbf{dm}}
      \left(x\right)\right)\right)\right) &\equiv& 
\opr{tr}{\Big (}
{\opr{{\tilde O}_\mathbf{dm}}}
(P({\opr{{\tilde I}_\mathbf{dm}}}(x))){\Big )} \nonumber\\ &>& 1 - (p+\alpha)~,
\end{eqnarray}
where ${\opr{{\tilde I}_\mathbf{dm}}}\equiv
{\mathcal{M}_{{\{\mathrm{l}\rightarrow\mathrm{c}\}}}}\circ{\opr{I_\mathbf{dm}}}$,
${\opr{{\tilde O}_\mathbf{dm}}}\equiv
{\opr{O_\mathbf{dm}}}\circ {\mathcal{M}_{{\{\mathrm{c}\rightarrow\mathrm{l}\}}}}$.
We have thus established eq.(\ref{des-approximation-cond}) with $p^\prime \equiv p + \alpha$.  
Thus the total probability of failure that the overall computation yields the correct 
{\em final} result 
is bounded by the sum of two terms: (1) an upper bound on the 
failure probability for the overall computation when the 
quantum computation is perfectly implemented without errors, and (2) an upper bound on the implementation inaccuracy of the actual quantum computation implemented by a practical device.  
This result is both intuitively simple and technically subtle.  It is not surprising that 
the total failure probability 
for the overall computation to yield the correct final result 
is bounded by quantities that describe the two sources of error. 
However, it is not immediately obvious that these quantities should combine in such a simple 
way.  On the one hand,  
the quantities $p$ and $p^\prime$ represent bounds on 
probabilities for obtaining certain outcomes from a measurement of the final quantum state.  
On the other hand, 
the quantity $\alpha$ is 
a bound, not on a probability, but on the normed difference between the states resulting 
from idealized and actual quantum computations, respectively.  

Diagram (\ref{nearly-commutative-diagram-1}) and eq.(\ref{approximation-cond-for-devices}) 
relate the actual implementation $P$ of the 
quantum computation to the intended calculation of $F$. They state the criterion for
a realization, $P$, of a quantum computer, operating 
fault-tolerantly in the presence of residual errors, to correctly implement an instance of
the overall computation.  If $p^\prime$ is sufficiently small a majority vote algorithm 
will successfully identify the correct outcome $F(x)$.  For example, in the case $Y$ is binary, 
the 
calculation of $F$ succeeds by majority voting if $p^\prime <1/2$.

Note that in the idealized limit in which error correction {\em perfectly} and {\em permanently} 
removes all residual errors ({\em i.e.}, in the limit $\alpha = 0$), our result reduces to the 
corresponding result of the Kitaev model ({\em {i.e.}}, eq.(\ref{approximation-cond-for-devices}) reduces to eq.(\ref{approximation-cond})).

\section{Applications to Practical Specifications for Fault Tolerance}

\subsection{General Result Relating Fault Tolerance Theory to the Overall Computation}

The above analysis, in which the constraints of fault tolerance are explicitly combined with those of the Kitaev model, has important practical applications to the specification of error tolerances for quantum computers.  The designer of a quantum computer wishes to achieve some upper bound $p^\prime$ on the probability that 
the final measurement of the final quantum state produced by the quantum computation {\em fails} to yield the correct final result for the overall computation.  
As discussed above, there is some bound, $p$, on the inherent probability that the overall computation 
will fail even if the quantum computation is perfectly implemented.  As discussed above, this is
due to the intrinsic nondeterminism inherent in the final measurement of the final quantum state resulting from the execution of the quantum computation, and it is an abstract property of the algorithm itself. Our result shows that the implementation of the algorithm by a realistic, {\it i.e.} noisy, quantum computer will meet the 
designer's success criterion provided that the implementation inaccuracy  satisfies 
$\|\mathcal{M}_{\{\mathrm{c}\rightarrow\mathrm{l}\}}(P\cdot(\mathcal{M}_{\{\mathrm{l}\rightarrow\mathrm{c}\}}(\rho))) -
G(\rho )\|_1 \leq p^\prime - p$.  
Since the ideal failure probability bound, $p$, is a characteristic 
of the ideal quantum algorithm itself, 
this result effectively apportions the allowable noise in a quantum 
computation between a component, $p$, due to the inherent quantum mechanical indeterminacy associated to the measurement of the final quantum state, and a component, $\alpha$, associated to the dynamics of the quantum computation itself, which includes other sources of noise, such as decoherence.  
Eq.(\ref{QC}) thus provides a success criterion for the design and implementation of a fault-tolerant quantum computation that, upon measurement of the final quantum state of which, produces the {\em correct final result} for the overall computation.  

As an example of how our result can be applied in order to achieve practical constraints on 
design parameters, suppose we wish to build a fault tolerant quantum computer which provides 
the correct solution to some specific problem ({\it e.g.} factoring a large number) 
with probability $1-\hat{p}$ or better.  That is, $\hat{p}$ is the required upper bound on 
the probability that the overall computation produces the wrong final result.  
From eq.(\ref{approximation-cond-for-devices}) 
we infer that the failure probability of the overall computation {\em as implemented} is 
bounded by  
$p+\alpha$.  We therefore require
\begin{equation}
\label{basic_req}
p+\alpha = \hat{p}~.
\end{equation}
The quantity $p$ is an intrinsic characteristic  
of the quantum algorithm, and can always, in principle, be determined. We therefore require
\begin{equation}
\alpha = \hat{p} - p~,
\end{equation}
which is a bound on the implementation inaccuracy of the quantum computation,  
sufficient to meet the prescribed success requirement for the overall computation.  
From eq.(\ref{QC}), we see that this 
criterion is met if
\begin{equation}
\label{derived_req}
\|\mathcal{M}_{\{\mathrm{c}\rightarrow\mathrm{l}\}}(P\cdot(\mathcal{M}_{\{\mathrm{l}\rightarrow\mathrm{c}\}}(\rho))) -
G(\rho )\|_1 \leq \hat{p} - p
\end{equation}
for all $\rho$.  

Techniques of fault tolerance theory can be used to determine the probability 
$\epsilon_{QC}$ that the quantum computation fails to produce the desired final 
quantum state.  
We therefore write the final state of the {\em logical} qubits resulting from the 
actual, practically-implemented, quantum computation as
\begin{equation}
\label{errstate}
\mathcal{M}_{\{\mathrm{c}\rightarrow\mathrm{l}\}}(P\cdot(\mathcal{M}_{\{\mathrm{l}\rightarrow\mathrm{c}\}}(\rho))) = 
(1-\epsilon_{QC}) G(\rho ) + \epsilon_{QC} \rho_{err}~,
\end{equation}
where $G(\rho)$ would be the result of an idealized quantum computation, $\rho_{err}$ arises 
from errors, and the {\em entire right hand side} 
of eq. (\ref{errstate}) is the state that results when errors occur.   
Eq. (\ref{derived_req}) then becomes 
\begin{equation}
\label{derived_req2}
\| \epsilon_{QC} \lbrack \rho_{err} - G(\rho) \rbrack \|_1 = 
\epsilon_{QC} \| \rho_{err} - G(\rho)  \|_1
\leq \hat{p} - p~.
\end{equation}
Since, 
\begin{equation}
\| \rho_{err} - G(\rho) \|_1 \leq 2~, 
\end{equation}
eqs. (\ref{derived_req2}), 
(\ref{derived_req}), and therefore (\ref{basic_req})  will be satisfied provided 
\begin{equation}
\label{general}
\epsilon_{QC}  
\leq \frac{1}{2} \left( \hat{p} - p \right) ~.  
\end{equation}
This is a new, quite general result relating the fault tolerance theoretic parameter 
$\epsilon_{QC}$ to the constraints we have derived, which ensure that the overall 
computation yield the correct final result with some prescribed success probability. 
This general inequality at once combines the constraints dictated by the QCC, which apply to the dynamics of the quantum computation, with those coming from the Kitaev model, which apply to the measurement, in terms of the numerical parameters of standard fault tolerance theory, which apply only to gate failure probabilities.

\subsection{Direct Example: Calculation of the Concatenation Level Function}

\begin{figure}
\begin{center}
\includegraphics[width = 8cm]{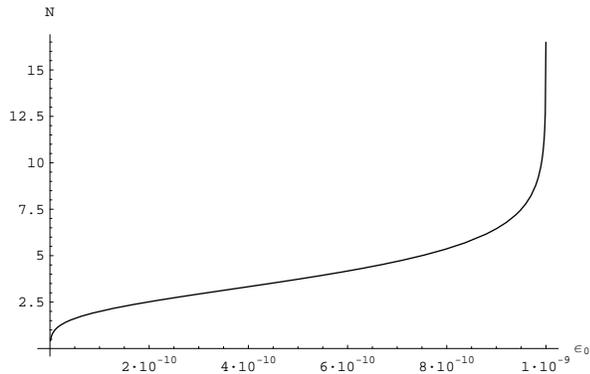}
\caption{\label{concatplot}
Levels of concatenation, $N$, required to meet the performance criterion as a function of elementary gate error probability, $\epsilon_0$. The performance criterion is 
to achieve the correct final result for the overall computation (not simply the desired final quantum state) with some prescribed success probability. In this example the success probability ($\hat{p}$) 
is prescribed to be 0.6. The bound on error probability associated to the measurement following ideal quantum computation ($p$) is taken to be 0.2. The number of gates ($\mathcal{N}$) is $10^{12}$, and the error threshold ($\epsilon_{th}$) is taken to be $10^{-9}$.}
\end{center}
\end{figure}

As an example of how our general result, Eq.(\ref{general}), 
can be applied to the specification of practical 
fault-tolerance design 
parameters, we consider the fault tolerance approach 
described in \cite{Preskill}, in which a concatenated quantum error correcting code is 
applied 
so as to reduce the failure probability at each level of concatenation.  
The failure probability for a single logical gate scales roughly as \cite{Preskill} 
\begin{equation}
\epsilon_N \simeq \epsilon_{th} \left( \frac{\epsilon_0}{\epsilon_{th}} \right)^{2^N}~,
\end{equation}
where $\epsilon_0$ is the probability of failure for elementary gates, $\epsilon_{th}$ is 
the fault tolerance threshold, and $\epsilon_N$ is the failure probability of the gate 
at the $N$th level of concatenation.   
For simplicity, in this example we assume that this relation is exact, that it applies equally to 
all gates, and that no other sources of error are present.  If the quantum 
computation is comprised of $\mathcal{N}$ logical gates, then the failure probability 
for the quantum computation to yield the desired final quantum state is
\begin{equation}
\epsilon_{QC} \simeq \mathcal{N} \epsilon_{th} \left( \frac{\epsilon_0}{\epsilon_{th}} \right)^{2^N}~.
\end{equation}
Thus, in order to ensure that we obtain the correct final result to the overall computation 
with sufficient probability of success, we make use of our general result in eq.(\ref{general}) to require that 
\begin{equation}
\epsilon_{QC} \simeq \mathcal{N} \epsilon_{th} \left( \frac{\epsilon_0}{\epsilon_{th}} \right)^{2^N} 
\leq \frac{1}{2} \left( \hat{p} - p \right) 
\end{equation}
is satisfied.  
From this we infer that the number of levels of concatenation sufficient to 
guarantee that the overall computation performs as required is given by:
\begin{equation}
N \gtrsim \log_2 
\frac{\ln \frac{2 \mathcal{N} \epsilon_{th}}{\hat{p}-p}}
     {\ln \frac{\epsilon_{th}}{\epsilon_0}} ~.
\end{equation}
This is a sufficient, not a necessary, condition.  Alternatively, for a given level of concatenation, we could just as well derive a requirement on the error probability 
$\epsilon_0$ for elementary gates.  In other words, this result enables us to explore the 
tradeoff between adding 
additional levels of concatenation and improving the performance of elementary gates 
in order to achieve a given 
performance criterion in terms of the reliability of the final result.  
An example of such a tradeoff curve is shown in Figure \ref{concatplot}. We stress that 
the points on the curve in Figure \ref{concatplot} do not merely represent the amount 
of concatenation needed in order that the {\em quantum computation} produce a particular final 
quantum state, but rather, fully incorporate the conditions that will ensure a successful 
{\em overall computation}.  

We emphasize that the role of standard fault tolerance theory in our calculation 
is solely to express 
the error probability for the quantum computation in terms of elementary gate errors.  
Standard fault tolerance theory does not provide estimates of (nor bounds on) the implementation inaccuracy as 
defined in eq.(\ref{QC}), and, consequently, would not have allowed us to calculate 
the number of levels of concatenation required in order to ensure that the overall computation produces the correct final result with the prescribed probability of success.  


\section{Conclusion}

In this paper we have introduced a methodology for determining design parameters for a quantum 
computer, the values of which ensure 
that an overall computation of interest, comprised of an initial purely 
quantum computation followed by a measurement of the final quantum state produced by the 
quantum computation, yields a correct {\em final result} with some prescribed probability of 
success, as opposed to merely ensuring that the desired {\em final quantum state} 
is obtained. Thus, our method enables the determination 
of {\em complete} quantum computational resource requirements associated to the actual solution 
of practical problems

Our method fully accounts for two sources of error that may affect the final 
result 
of the overall computation: (1) errors which arise in the course 
of the quantum computation itself, and (2) errors due to the indeterminacy intrinsic to the final 
measurement which follows the quantum computation. Standard fault tolerance 
theory  
addresses the first type of 
error, but does not consider the second.  Kitaev's model addresses the second type of error, but 
not the first.  
We have extended both standard fault tolerance theory and Kitaev's model,  
and have combined them, in order to yield quantitative results 
that reveal the interplay between the 
two.  Although the analysis in this paper has been presented in the framework of the 
circuit paradigm for quantum computing, it is straightforward to apply our results to 
other paradigms, including the cluster state and adiabatic 
paradigms~\cite{ghwprog}.  

As a specific example of the practical application of our approach, 
we have explicitly calculated the number of levels of error correction 
concatenation needed to achieve a correct final result for the overall computation with 
some prescribed success probability. Extensions of the current calculation will include associating different failure probabilities to different gates, as well as considering additional refinements dictated by imposing the QCC on the overall dynamics of the quantum computer~\cite{ghwprog}.

\begin{acknowledgements} 
\noindent

The authors thank F. Javier Thayer for his contributions to the early phases of this research. They also thank Stephen P. Pappas and Anthony Donadio for their input. This research was supported under MITRE Technology Program Grant 07MSR205.
\end{acknowledgements}
\newpage



\begin{thebibliography}{000}
\bibitem{f1}
For simplicity we do not consider errors in the 
preparation of the initial quantum state in the present paper.
\end{thebibliography}
\end{document}